# Functional optoacoustic neuro-tomography (FONT) for whole-brain monitoring of calcium indicators

Gali Sela[1], Antonella Lauri[1,2], X. Luís Deán-Ben[1], Moritz Kneipp[1,3], Vasilis Ntziachristos[1,3], Shy Shoham[4,*], Gil G. Westmeyer[1,2,3,*], and Daniel Razansky[1,3,*]

[1]*Institute for Biological and Medical Imaging (IBMI), Helmholtz Center Munich, Neuherberg, Germany*
[2]*Institute of Developmental Genetics, Helmholtz Center Munich, Neuherberg, Germany*
[3]*Department of Medicine, Technische Univeristät München, Munich, Germany*
[4]*Department of Biomedical Engineering, Technion – Israel Institute of Technology, Haifa, Israel*

**Non-invasive observation of spatiotemporal neural activity of large neural populations distributed over entire brains is a longstanding goal of neuroscience. We developed a real-time volumetric and multispectral optoacoustic tomography platform for imaging of neural activation deep in scattering brains. The system can record 100 volumetric frames per second across a 200mm³ field of view and spatial resolutions below 70µm. Experiments performed in immobilized and freely swimming larvae and in adult zebrafish brains demonstrate, for the first time, the fundamental ability to optoacoustically track neural calcium dynamics in animals labeled with genetically encoded calcium indicator GCaMP5G, while overcoming the longstanding penetration barrier of optical imaging in scattering brains. The newly developed platform offers unprecedented capabilities for functional whole-brain observations of fast calcium dynamics; in combination with optoacoustics' well-established capacity in resolving vascular hemodynamics, it could open new vistas in the study of neural activity and neurovascular coupling in health and disease.**

Neuronal activation occurs concurrently or in a highly coordinated fashion in different areas across the nervous system, reflecting functional interconnection between specialized neuronal sub-circuits. Imaging neuronal activation with high temporal and spatial resolution over an entire intact brain, including deep and normally inaccessible areas, could thus play a critical role in the attempt to decipher the fundamental operating principles underlying neural circuit activity. Major efforts are underway to advance the ability to optically image the activity of large, distributed neural populations[1,2]. Recently, these efforts have led to whole-brain activity imaging in transparent organisms using single[3] and multi-photon[4] light-sheet and light-field[5] microscopy. However, these approaches are unable to resolve deep neural activity in intact *scattering* brains where current state-of-the-art optical imaging strategies based on rapidly scanning multiphoton microscopy are limited to volumes below 1mm³ due to limited penetration depth and field of view (FOV).

Conversely, optoacoustic imaging, an extremely powerful approach with demonstrated capacity for centimeter scale penetration into highly scattering tissues[6], can potentially provide functional neural imaging beyond the limit of current optical imaging technologies. To date, functional optoacoustic brain imaging mainly focused on probing hemodynamics and blood oxygenation variations[7–9], slow and delayed processes that only indirectly reflect neural activation. In contrast, modern approaches towards imaging neural activation are largely based on measuring calcium dynamics, which provides a much more direct correlate of neural activity. Calcium dynamics are conventionally tracked using fluorescent sensors, which change their fluorescence output because of a change in their extinction coefficient, fluorescence quantum yield or both as a function of intracellular calcium concentrations[10,11]. Recent advances in the field of genetically encoded calcium indicators (GECIs) have provided a variety of genetically modified reporter animal models with calcium indicators in specific neuronal structures[12–14], which in some cases can give rise to signals down to the level of single action potentials[13,15].

We hypothesized that some members of the GCaMP5G family of fluorescent GECIs could provide a suitable calcium dependent optoacoustic contrast due to their strong calcium-dependent absorption changes in the presence of calcium[12]. To test this hypothesis, we selected the zebrafish reporter line *HuC:GCAMP5G* that exhibits expression of the calcium sensor in a large fraction of neurons to obtain high-resolution optoacoustic measurements from larvae and isolated adult brains.

To enable tomographic imaging of the whole scattering brain in this model system with high spatiotemporal resolution, we developed a new generation functional optoacoustic neuro-tomography (FONT) system that can simultaneously monitor calcium activity in all five dimensions, *i.e.* space, time and wavelength (spectrum) (Fig. 1A). The experimental setup and accompanying image reconstruction algorithms enable real-time rendering of ~200mm³ volumes at a frame rate of 100Hz; this temporal resolution is high enough to allow tracking of freely moving organisms (e.g., a freely swimming 6 day old wild type zebrafish larva, supplementary video 1). This major improvement over previous state-of-the-art implementations[16] further enables the acquisition of over $10^6$ informative voxels over a 6 · 6 · 6 mm FOV with spatial resolution down to 52µm and 71µm in the axial and lateral directions, respectively (Fig. 1B). The laser also has fast wavelength tuning capability, enabling the fast acquisition of spectroscopic information regarding chromophores of interest. To explore the fundamental spectral signature of our experimental system, 6 days old HuC:GCaMP5G zebrafish larvae were imaged using two distinct wavelengths: 488nm where GCaMP5G has high absorption and 530nm where GCaMP5G absorption is close to zero (Fig 1C). The larva's eyes, with their characteristically highly absorbing melanin pigmentation are readily visible in both optoacoustic images, whereas tissues expressing GCaMP5G do not yield a strong signal at 530 nm but provide clear contrast at 488nm, where the protein is highly absorbing (Fig. 1D). Optoacoustic measurement of calcium-related dynamic changes in GCaMP5G absorption were compared with planar fluorescence by coupling the system to a fast sCMOS camera.

\* Correspondence to S. S (sshoham@bm.technion.ac.il), G. G. W. (gil.westmeyer@tum.de) or D. R. (dr@tum.de)

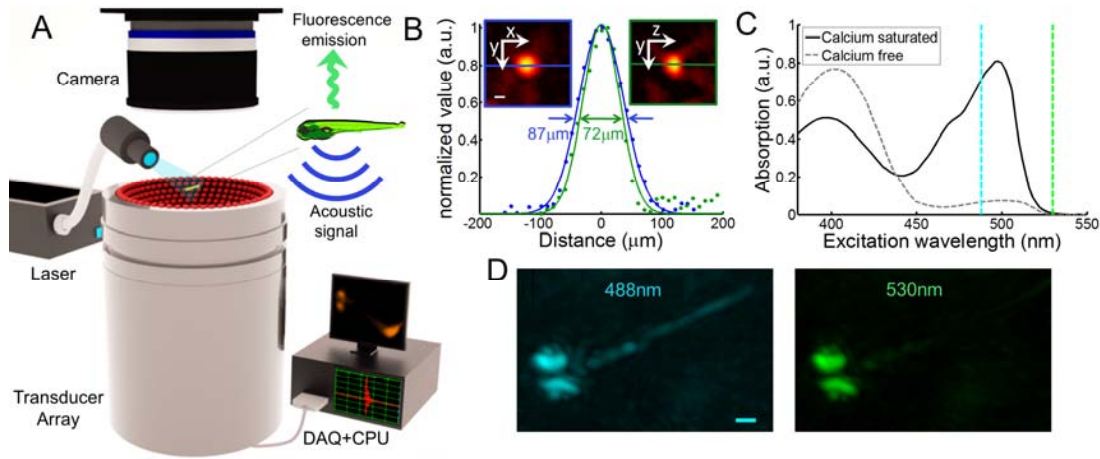

**Figure 1. FONT setup.** (**A**) The imaged sample is placed in the vicinity of the geometrical center of a custom-made spherical ultrasound detection array. Light illumination of the entire object is provided via an optical fiber bundle. The generated optoacoustic signals are acquired by the array transducer while the induced fluorescence is simultaneously recorded by the high speed sCMOS camera. (**B**) Characterization of the spatial resolution of the volumetric optoacoustic imaging system was done by analyzing 3D optoacoustic images from an absorbing sphere (shown as insets with scale bar of 50µm). Considering the 50µm diameter of the microsphere, the depicted one-dimensional signal profiles correspond to lateral and axial spatial resolution of 71µm and 52µm, respectively. (**C**) Spectral dependence of the absorbance on calcium-bound versus calcium free GCaMP5G (reproduced from [12]). (**D**) Optoacoustic images of a 6 days old *HuC:GCAMP5G* larva acquired at 488nm and 530nm (maximal intensity projection (MIP) views through volumetric reconstructions are shown). While the pigmented eyes are strongly absorbing at both wavelengths, GCaMP5G is hardly absorbing at 530nm, thus neural tissue is much more visible at 488nm (scale bar - 250µm).

To elicit and image strong neural activation we exposed immobilized HuC:GCaMP5G zebrafish larvae (n=5) to a neuroactivating agent (Pentylenetetrazole, PTZ) that induces fast ictal-like spikes in the larvae's nervous system and changes in their swimming behavior[17], most likely by interfering with GABAergic signaling[18]. PTZ exposure caused robust calcium waves propagating from the larva's posterior (site of drug injection) to the anterior parts of its spinal cord (Fig. 2). Strong correlations between the simultaneously acquired optoacoustic and fluorescent signals were observed during evoked calcium transients both at the tip ($R^2=0.98$) and midtail ($R^2=0.97$) regions, providing confirmation that GCaMP5G calcium sensing can be read out via optoacoustics (Fig. 2 and supplementary video 2).

Next, we examined FONT's imaging performance in isolated brains of *adult* HuC:GCaMP5G zebrafish (n=4), measuring 2-3 mm on their short axis. The system was found to provide high-quality time-resolved 3D reconstructions across these *highly scattering* brains (Fig. 3); note that only the forebrain and part of the optic tectum were effectively illuminated and thus visualized. Five volumes of interest (VOIs) were then selected for analysis of the dynamics of GCaMP5G using time series of both the optoacoustic (absorption) and the planar fluorescence signals during neural activation (Fig. 3B). According to the optoacoustic signal traces, activation patterns associated with high contrast calcium-related changes occurred mainly in deep brain areas with maximal contrast $\Delta OA/OA_0=8.5$, while voxels close to the surface actually showed a slight decrease in activity (blue VOI). However, as compared to true 3D information provided by the tomographic optoacoustic reconstructions, whose spatial resolution is not affected by the intense light scattering, the planar fluorescence lacks optical sectioning thus may result in wrong conclusions based on smeared sub-surface information averaged over large volumes[19]. The signal traces from the gray VOI placed over a superficial part of the optic nerve show a similar overall trend in both the fluorescence and optoacoustic modality, although the optoacoustic signal shows some faster signal fluctuations (Fig. 3B and supplementary video 3). However, signal changes can be detected in the pink VOI by optoacoustics that are absent in the planar fluorescence images. Similarly, the time course averaged over the yellow VOI placed deeper in the brain shows signal changes that are not detected by the fluorescence read out from an approximately corresponding planar region. This simultaneous dual-mode imaging demonstrates that epi-fluorescence fails to faithfully identify the calcium fluxes that optoacoustics localized deep inside scattering brain.

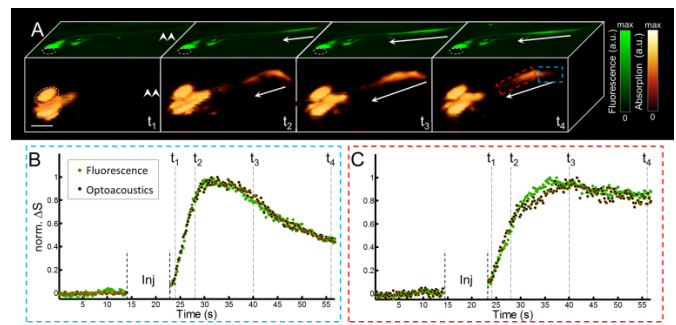

**Figure 2. Imaging of neuronal activation in *HuC:GCAMP5g* zebrafish larvae.** (**A**) Planar epi-fluorescence (green) and 3D optoacoustic (absorption contrast - orange) images of a 6 day old larvae at 4 time points after injection of neurostimulant show a similar dynamics of the signals (scale bar - 500µm). The arrows show direction of the activation as it progresses from the posterior (site of injection marked by a double arrow in the left frame) to the anterior part of the tail. Eyes are marked with a dotted line (**B**) Traces of the fluorescence and optoacoustic signal changes in the posterior region of spinal cord (marked by red rectangle in panel A). (**C**) Corresponding traces in the anterior region (marked by the blue rectangle). The time points shown in panel A are marked by vertical dotted lines. Since both the background fluorescence and optoacoustic signals were close or below the noise level, the changes in the signals (ΔS) were normalized to unity instead of being divided by the respective resting signals. The injection phase caused image artifacts and is therefore excluded from the graphs.

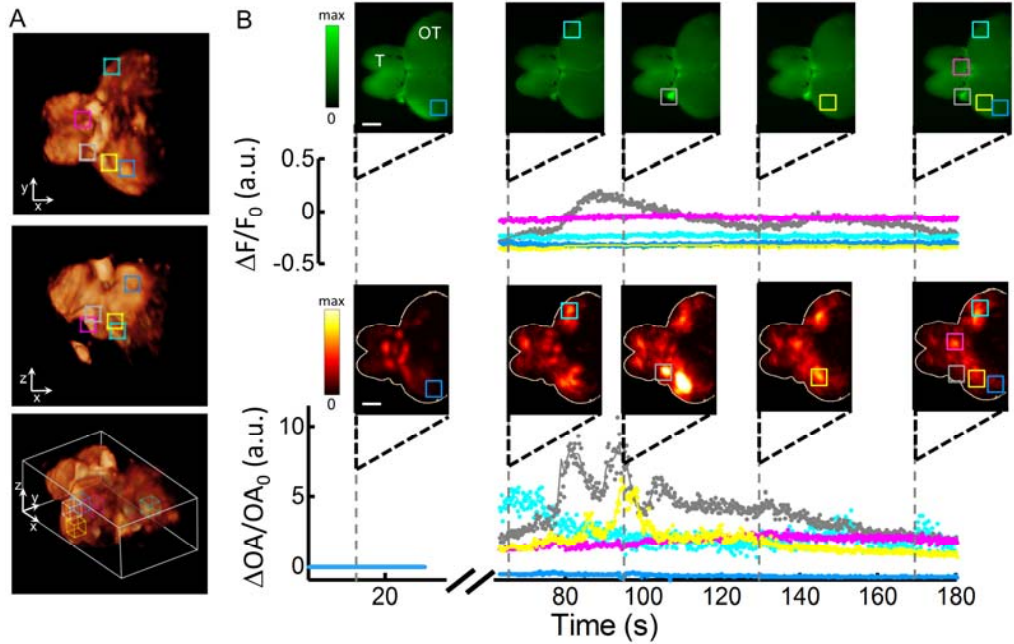

**Figure 3. Activity in isolated scattering brains.** (**A**) Typical 3D optoacoustic image acquired from a highly scattering brain of an adult fish in its resting state (two side views and one isometric view of the 3D reconstruction are shown). The telencephalon (T) and most of the optic tectum (OT) are clearly visible. Five 300µm x 300µm x 300µm VOIs were chosen at different locations and depths within the brain. (**B**) Traces of the fluorescence (top) and optoacoustic (bottom) signal changes are shows for the five regions in corresponding colors (all signal changes are normalized to the resting signal levels). Note that the optoacoustic traces are calculated over volumes whereas the fluorescent signals are calculated over the roughly corresponding planar areas. Snapshots acquired at 5 different time points before and after introduction of the neurostimulant are shown (scale bar - 500 µm). The injection phase caused image artifacts and is therefore excluded from the graphs.

Finally, we examined whether the system's exceptionally large FOV opens up the capability to non-invasively measure neural activity while simultaneously monitoring ongoing natural behavior. Indeed, one of neuroscience's major challenges is to study neural processing during unrestrained motion, motivating the successful recent introduction of a number of experimental paradigms for studying behaving zebrafish larvae using e.g. bioluminescence imaging[20] as well as light sheet[21] and light field[5] microscopies (note however that these strategies cannot provide volumetric information in scattering brains). We therefore tested whether our optoacoustic system enables high-speed volumetric imaging in larvae that were allowed to swim freely in ~0.5cm$^2$ chamber and then exposed to the neurostimulant PTZ - leading to rapid movements followed by long resting periods. The optoacoustically-recorded responses indeed revealed an increase up to $\Delta OA/OA_0=1.8$ in calcium signal (see arrows in Fig. 4), just before the fish moves.

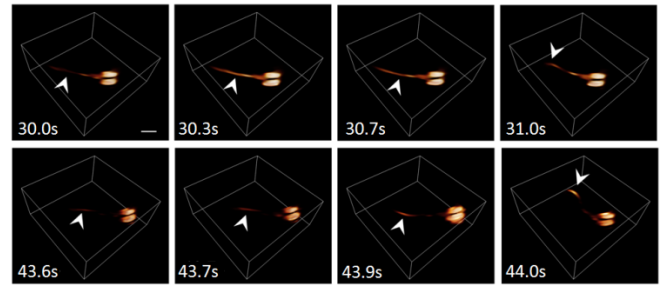

**Figure 4. Activation in freely swimming larvae.** Two separate activation events, as captured by volumetric optoacoustic tomography, are shown. Following injection of the neurostimulating agent at approximately $t$=0, the larva occasionally stops swimming while experiencing a surge of activation through its tail (the arrows point to the location of optoacoustic signal increase) before it starts moving promptly to a new position (notice movement of the tail in the two rightmost frames). Scale bar - 500µm.

In summary, we demonstrated a novel, state-of-the-art optoacoustic imaging platform for direct imaging of spatio-temporal neural activity across an entire light-scattering brain while maintaining similar values of spatial resolution at highly scalable depths. Our study is also, to the best of our knowledge, the first to examine the optoacoustic signature of modern GECIs, showing that the strong changes in GCaMP5G fluorescence are directly related to their optoacoustic signature. Since FONT uses a widely-tunable nanosecond OPO laser technology, it can be conveniently tuned to work with a large array of other functional probes, including for instance the newer generation of GCaMP6 probes [13] and red shifted probes like RGECO[22], or future sensors optimized for *in vivo* deep tissue optoacoustic detection providing high extinction coefficient changes in the near-infrared window as well as lower quantum yield. Furthermore, the ability to simultaneously track the movement and neural activation of living unrestrained organisms could form a basis for behavioral studies not currently possible with other neuroimaging techniques. While the current spatial resolution does not allow to distinguish individual cells, future generations of the described tomographic approach could utilize higher-frequency transducer technology and/or super-resolution strategies for fast functional observations at the cellular-scale. In addition, the high sensitivity of optoacoustics to a variety of intrinsic absorption tissue contrasts, most prominently the oxy- and de-oxy hemoglobin, is well established[23,24]. This may provide highly complementary information to the functional calcium imaging and thus enhance the amount of spectrally- and time-resolved volumetric information available for the five dimensional optoacoustic studies looking at coupling between the vascular changes and nervous system in more complex animal models. While we recently demonstrated high-resolution imaging of near-infrared fluorescent proteins (iRFP) in mouse brain in-vivo using multispectral optoacoustic tomography[25], development of similar calcium sensitive indicators in the near-infrared is expected to greatly expedite translation of our technology into mammalian brains.

## Acknowledgements

The authors acknowledge grant support from the European Research Council *ERC-2010-StG-260991* (D. R.) and ERC-2012-StG _20111109 (A. L. and G.G.W.), the German-Israeli Foundation (GIF) for Scientific Research and Development (S. S. and D. R.), the Helmholtz Association of German Research Centers and the Technische Universität München (G. G. W.). The authors acknowledge technical assistance from M. Sadasivam, A. Lin, and L. Ding.

## References

1. Adesnik, H., Waller, L. & Shoham, S. Optics on the Brain: OSA's Mulitphoton and Patterned Optogenetics Incubator. *Opt. Photonics News* **25,** 42–49 (2014).
2. Dana, H. *et al.* Hybrid multiphoton volumetric functional imaging of large-scale bioengineered neuronal networks. *Nat. Commun.* **5,** (2014).
3. Ahrens, M. B., Orger, M. B., Robson, D. N., Li, J. M. & Keller, P. J. Whole-brain functional imaging at cellular resolution using light-sheet microscopy. *Nat. Methods* **10,** 413–20 (2013).
4. Schrödel, T., Prevedel, R., Aumayr, K., Zimmer, M. & Vaziri, A. Brain-wide 3D imaging of neuronal activity in Caenorhabditis elegans with sculpted light. *Nat. Methods* **10,** 1013–20 (2013).
5. Prevedel, R. *et al.* Simultaneous whole-animal 3D imaging of neuronal activity using light-field microscopy. *Nat. Methods* **11,** 727–30 (2014).
6. Wang, L. V & Hu, S. Photoacoustic tomography: in vivo imaging from organelles to organs. *Science* **335,** 1458–62 (2012).
7. Nasiriavanaki, M. *et al.* High-resolution photoacoustic tomography of resting-state functional connectivity in the mouse brain. *Proc. Natl. Acad. Sci. U. S. A.* **111,** 21–6 (2014).
8. Yao, J. *et al.* Noninvasive photoacoustic computed tomography of mouse brain metabolism in vivo. *Neuroimage* **64,** 257–66 (2013).
9. Yao, L., Xi, L. & Jiang, H. Photoacoustic computed microscopy. *Sci. Rep.* **4,** 4960 (2014).
10. Takahashi, A., Camacho, P., Lechleiter, J. D. & Herman, B. Measurement of intracellular calcium. *Physiol. Rev.* **79,** 1089–125 (1999).
11. Wilms, C. D. & Eilers, J. Photo-physical properties of Ca2+-indicator dyes suitable for two-photon fluorescence-lifetime recordings. *J. Microsc.* **225,** 209–13 (2007).
12. Akerboom, J. *et al.* Optimization of a GCaMP calcium indicator for neural activity imaging. *J. Neurosci.* **32,** 13819–40 (2012).
13. Chen, T. W. *et al.* Ultrasensitive fluorescent proteins for imaging neuronal activity. *Nature* **499,** 295–300 (2013).
14. Hires, S. A., Tian, L. & Looger, L. L. Reporting neural activity with genetically encoded calcium indicators. *Brain Cell Biol.* **36,** 69–86 (2008).
15. Wilms, C. D. & Häusser, M. Twitching towards the ideal calcium sensor. *Nat. Methods* **11,** 139–140 (2014).
16. Deán-Ben, X. L. & Razansky, D. Portable spherical array probe for volumetric real-time optoacoustic imaging at centimeter-scale depths. *Opt. Express* **21,** 28062–28071 (2013).
17. Baraban, S. C., Taylor, M. R., Castro, P. a & Baier, H. Pentylenetetrazole induced changes in zebrafish behavior, neural activity and c-fos expression. *Neuroscience* **131,** 759–68 (2005).
18. Leweke, F., Louvel, J., Rausche, G. & Heinemann, U. Effects of pentetrazol on neuronal activity and on extracellular calcium concentration in rat hippocampal slices. *Epilepsy Res.* **6,** 187–198 (1990).
19. Buehler, A. *et al.* High resolution tumor targeting in living mice by means of multispectral optoacoustic tomography. *EJNMMI Res.* **2,** 14 (2012).
20. Naumann, E. A., Kampff, A. R., Prober, D. A., Schier, A. F. & Engert, F. Monitoring neural activity with bioluminescence during natural behavior. *Nat. Neurosci.* **13,** 513–20 (2010).
21. Vladimirov, N. *et al.* Light-sheet functional imaging in fictively behaving zebrafish. *Nat. Methods* **11,** 1–2 (2014).
22. Akerboom, J. *et al.* Genetically encoded calcium indicators for multi-color neural activity imaging and combination with optogenetics. *Front. Mol. Neurosci.* **6,** 2 (2013).
23. Beard, P. Biomedical photoacoustic imaging. *Interface Focus* **1,** 602–631 (2011).
24. Xu, M. & Wang, L. V. Photoacoustic imaging in biomedicine. *Rev. Sci. Instrum.* **77,** 041101 (2006).
25. Deliolanis, N. C. *et al.* Deep-tissue reporter-gene imaging with fluorescence and optoacoustic tomography: a performance overview. *Mol. Imaging Biol.* **16,** 652–60 (2014).